\documentclass[12pt]{article}
\usepackage{graphicx}

\newcommand\pubnumber{NOBUGS2002/017}
\newcommand\pubdate{\today}

\def\frm2{ZWE FRM-II\\
Technische Universit\"at M\"unchen\\
Lichtenbergstr. 1\\
D-85748 Garching, F.R.G.\\
jens.krueger@frm2.tu-muenchen.de\\
}

\def\Title#1{\begin{center} {\Large #1 } \end{center}}
\def\Author#1{\begin{center}{ \sc #1} \end{center}}
\def\Address#1{\begin{center}{ #1} \end{center}}

\newcommand\pubblock{\rightline{\begin{tabular}{l} \pubnumber\\
         \pubdate  \end{tabular}}}
\newenvironment{Abstract}{\begin{quotation}  }{\end{quotation}}
\newenvironment{Presented}{\begin{quotation} \begin{center} 
             PRESENTED AT\end{center}\bigskip 
      \begin{center}\begin{large}}{\end{large}\end{center} \end{quotation}}

\def\beq{\begin{equation}}
\def\eeq#1{\label{#1}\end{equation}}
\def\eeqn{\end{equation}}

\def\beqa{\begin{eqnarray}}
\def\eeqa#1{\label{#1}\end{eqnarray}}
\def\eeqan{\end{eqnarray}}

\let\bar=\overbar

\def\Dslash{\not{\hbox{\kern-4pt $D$}}}
\def\dslash{\not{\hbox{\kern-2pt $\del$}}}

\def\msb{{\bar{\ssstyle M \kern -1pt S}}}

\begin{document}
\begin{titlepage}
\pubblock
\vfill
\Title{{\tt autotools} for writing portable software}
\vfill
\Author{Jens Kr\"uger}
\Address{\frm2}
\vfill

\begin{Abstract}
Most of the scientific software is developed in the same manner: It starts at a 
lab and the scientist says "This part of software is only for my private use". But the software
grows and grows, and it comes the time, that other scientists want to use it, too. If they
are not working in the same lab and on the same machine, it leads to the problem how to
install this software on the new machine. 
Sometimes it is relatively easy to adopt the {\tt makefile} to the new conditions (compiler, linker, libraries), 
if it is well written. 
But mostly it is not done by changing some pathes. One has to change the compiler, linker, additional or 
standard libraries, and so on. I want to show how you avoid a lot of trouble, if you write software, which may be
used on different platforms.
\end{Abstract}
\vfill
\begin{Presented}
NOBUGS 2002\\
Gaithersburg, U.S.A,  November 04--06, 2002
\end{Presented}
\vfill
\end{titlepage}

\def\thefootnote{\fnsymbol{footnote}}
\setcounter{footnote}{0}

\section{Introduction}

If you start to write a software, you cannot imagine that anybody wants to use it, too. It's often that you 
start to write a part of software for your own use only. 

An example: You want to analyse your experimental data. Your data are stored in an electronic form, and you want
to find out the maximum count for each spectrum you measured. The first approach is to look at the files if they are stored
in a human readable format and find out the maximum by searching by hand. It is a stupid job. This may also been
done by a computer. It is foolish enough to do this job if you tell it, what is to be done. The next step is to write all
maxima for all files you analysed and draw a plot with these values. The written program helps you to analyse the data.
You tell this to one of your colleagues who wants to use it, too. But his machine is not same as yours. He has
a different operation system, compiler, etc.

To install this software on his machine you have to change your \texttt{Makefile} (which hopefully exists). Afterwards 
you try to create your
software on this machine. After some efforts you may finish this job. But actually you would like to have a system
doing this job for you.
You remembered that in some software packages you downloaded from the WWW they told you to install a
package by typing in only the following three commands:
\begin{verbatim}
user> ./configure
user> make
user> make install
\end{verbatim}
and the software is installed and useable. 

You wondered, where this mysterious {\tt configure} comes from and who wrotes this nearly incomprehensible 
{\tt Makefile}, since your own {\tt Makefile} look much simpler. The wizards who have done these jobs are: the {\tt autotools}.

This set of tools comes from the GNU world. It consists of
three parts: {\tt autoconf}, {\tt automake}, and {\tt libtool}. {\tt autoconf} is designed for checking the system (testing
compiler, linker, its options and so on), {\tt automake} for creating a makefile with automatically added
standard rules from a simple template file, and {\tt libtool} assists the developer during the creation of
statically and/or dynamically linked libraries. All three may be used alone, but the combination of
all of them is the most efficient way.

I will give an introduction of using the {\tt autotools} in combination. I will show, how 
you can write a software which is portable from one to another system.

\section{Example}
Let's consider the following situation. You have some files belonging to the project (some of these
files build a library):
\begin{verbatim}
main.c
func_a.h
func_a.c
func_b.h
func_b.c
\end{verbatim}

The file \texttt{main.c} contains the code for the main program whereas \texttt{func\_a.c} and \texttt{func\_b.c} contain the code for 
two functions \texttt{A()} and  \texttt{B()} .

The files \texttt{func\_a.h} and \texttt{func\_b.h} contain the declarations of functions \texttt{A()} and \texttt{B()}.

You want to generate a program \texttt{myprog} and a library \texttt{libmylib}. The library may be created as 
static as well as as shared library.

You have to write two files, \texttt{configure.ac} and \texttt{Makefile.am} for doing this job. As you can imagine the 
file \texttt{configure.ac} will contain some information for the configuration of the software and the \texttt{Makefile.am}
some information for building the software from the sources.

At first let's have a look at the \texttt{configure.ac} file.

\section{\texttt{configure.ac} file}

The \texttt{configure.ac} file starts with a line:
\begin{verbatim}
AC_INIT(myproject, 1.0.0, [myproject-bug@myproject.org])
\end{verbatim}

This lines tells the system: Generate a project named \texttt{myproject} with the version number \texttt{1.0.0}. The last
parameter gives the email address by which the developer of this project may be contacted. 

The meaning of the next three lines is the following:

\begin{verbatim}
AC_PREREQ(2.52)
AC_COPYRIGHT([(c) Jens Krueger])
AC_REVISION([1.0])
\end{verbatim}

the system requires an autoconf version at least 2.52, it defines the copyright message and defines the version of this file.

By the help of 
\begin{verbatim}
AC_CONFIG_SRCDIR(func_a.c)
\end{verbatim}
the system may verify that it works on the right system. The parameter of \texttt{AC\_CONFIG\_SRCDIR} should point to a file 
unique for this project. 


\begin{verbatim}
AM_INIT_AUTOMAKE([1.6])
\end{verbatim}
gives \texttt{autotools} the input that at least the version 1.6 of \texttt{automake} is required.

So up to this point you have configured a lot of things only for the \texttt{autotools} itself, but nothing for your own project 
with the exception of the first line. For compiling your project you need a C compiler. To tell \texttt{autotools}: "Search for a 
C compiler and linker and find out the right options" you use the following line.

\begin{verbatim}
AC_PROG_CC
\end{verbatim}

The next line checks the existence of \texttt{libtool} which is very helpful for creating libraries. 
\begin{verbatim}
AC_PROG_LIBTOOL
\end{verbatim}

An useful feature of \texttt{autotools} is the preselection of an installation path. As developer you may preselect a path 
for the installation of the software package. If the user does not overwrite this during the configuration run, the default
installation path will be used. The line
\begin{verbatim}
AC_DEFAULT_PREFIX(/usr/local)
\end{verbatim}
sets the default installation path. This may be overwritten during the configuration time with the option 
\texttt{--prefix=/new-installation-path}.

The last lines are very interesting for the project. These lines tell the \texttt{autotools} which files have to be created.
The \texttt{AC\_CONFIG\_FILES} macro defines the file names to be created. For each file given in this list, 
a file with the extension "\texttt{.in}" has to exist. In your example you need a file \texttt{Makefile.in}. But where does it come from?
It comes from the \texttt{Makefile.am}. How this works will be explained in the next chapter. \texttt{AC\_OUTPUT} macro performs the 
creation of the files given by the \texttt{AC\_CONFIG\_FILES} macro.

\begin{verbatim}
AC_CONFIG_FILES(Makefile)
AC_OUTPUT
\end{verbatim}

\section{\texttt{Makefile.am} file}
As described in the previous chapter, the \texttt{Makefile.am} leads to a \texttt{Makefile} which can be used by the \texttt{make}
command. 

The \texttt{Makefile.am} contains the following lines:
\begin{verbatim}
bin_PROGRAMS = myprog
myprog_SOURCES = main.c
lib_LTLIBRARIES = libmylib.la
libmylib_la_SOURCES = func_a.c func_b.c
include_HEADERS = func_a.h func_b.h
\end{verbatim}

Now let's have a deeper look into the \texttt{Makefile.am}. The first line declares that you want to generate an executeable, which 
should be installed in the \texttt{\$prefix/bin} directory, called \texttt{myprog}. The following line gives information on the
source file(s) from which the \texttt{myprog} should be generated: \texttt{main.c}. If the \texttt{myprog} should also be generated 
from file \texttt{myprog.c} you have simply to add \texttt{myprog.c} to this line.

The third line tells you that you want to build a so-called libtool archive (\texttt{.la}). This is a library, but at this time you have not to 
decide whether you want to use static or shared libraries. This will be done at configuration time. The library will be installed in
the \texttt{\$prefix/lib} directory. 

The fourth line defines the sources from which the library has to be built up. Please pay attention to the fact, that the '.' has been
replaced by a underscore character.

The last line defines that the files \texttt{func\_a.h} and \texttt{func\_b.h} have to be installed in the \texttt{\$prefix/include}
directory.

\section{Preparing the system for a build}
Now you have finished the set-up of your project and you should start the creation of the \texttt{configure} file. 

By calling
\begin{verbatim}
user> libtoolize --copy --automake
\end{verbatim}
you tell the \texttt{libtool} package that you want to use it in cooperation with the \texttt{automake} package and it should copy all 
neccessary files into the local directory. Omitting the parameter \verb+--copy+ the \texttt{libtoolize} command only creates symbolic links
instead of copies of its administrative files.
 
The next step you have to do is to call the \texttt{aclocal} program, which extracts all macros used in the \texttt{configure.ac} and 
generates a file \texttt{aclocal.m4} later used by \texttt{autoconf}. 

The third call  
\begin{verbatim}
user> automake --copy --add-missing
\end{verbatim}
generates a template \texttt{Makefile.in} and installs some administrative files. Omitting the parameter \verb+--copy+ similarly to the 
\texttt{libtoolize} call leads to symbolic links instead of local copies of the admin files. You get some messages:
\begin{verbatim}
Makefile.am: required file `./NEWS' not found
Makefile.am: required file `./README' not found
Makefile.am: required file `./AUTHORS' not found
Makefile.am: required file `./ChangeLog' not found
\end{verbatim}
These missing files are required in the GNU world. It is a good idea to create these files by
\begin{verbatim}
user> touch NEWS README AUTHORS ChangeLog
\end{verbatim}
and fill them with useful information about your project. Additionally \texttt{automake} creates two other files \texttt{COPYING} and \texttt{INSTALL}
which contain the GNU Public License and the standard installation guide of the project. If you want to change the license or the installation 
guide simply change the content of these files. If you created them as symbolic links remove them and create regular files with the same name.

At last you have to call
\begin{verbatim}
user> autoconf
\end{verbatim}
which produces the \texttt{configure} script.

The preparations for generating a portable build process are finished. Now you configure your system by typing in: 
\begin{verbatim}
user> ./configure
\end{verbatim}

The output of this command tells you a lot of things, for example what kind of C compiler is installed and found on the system, whether the system 
may create static or shared libraries, whether they should be created, and so on. If the call does finish without any errors you may start your
build process by typing in
\begin{verbatim}
user> make
\end{verbatim}

If the build process ran successfully, you may install your package by typing:
\begin{verbatim}
user> make install
\end{verbatim}

The reverse process may be called by typing
\begin{verbatim}
user> make uninstall
\end{verbatim}
 
\section{Roll out}

If your package runs correctly you need a way for distributing your software in a tar'ed format. The simplest but not the best way
is to tar the complete directory, in which you built your software. The \texttt{autotools} help you with two targets in the 
\texttt{Makefile} which are generated automatically: \texttt{dist} and \texttt{distcheck}.

The \texttt{dist} target provides the creation of a \texttt{tar} file containing all necessary files for a build process on another machine.
The created \texttt{tar} file has to be installed on the target machine. But this way is not safe. A better way is to call the target \texttt{distcheck}.

This target creates a \texttt{tar} file, installs this in a temporary directory, calls the \texttt{configure} script and the \texttt{make} command
to build the software, to install, and to uninstall it. If the process was successful you get the message:
\begin{verbatim}
================================================
myproject-1.0.0.tar.gz is ready for distribution
================================================
\end{verbatim}

This message tells you that your software package distributed in the \\
\verb+myproject-1.0.0.tar.gz+ 
file may be installable
on other machines with different operating systems and architectures.

\section{Extending the configuration task}
At this point we have to make some remarks: Above a very simple example was discusssed to show you how to use the \texttt{autotools}. 
They are powerful and  provide a lot of macros for distinct tasks. So it is possible:
\begin{itemize}
\item to search for a special library
\item to search for a special function in a library
\item to look for a special header file on the target system
\item and, and, ...
\end{itemize}
\texttt{autotools} also give you the chance to create  a machine dependent config file which may be included in your sources. This controls
the compile process of the sources depending on the found header files, compilers, library versions and so on. 

Let's extend your project by some of these features. Assuming you want to use the NeXus library \cite{nexus} in your project 
some requirements have to be fulfilled. You have to install the HDF library \cite{hdf} on which NeXus is based and the NeXus library itself.

What do you have to do now? For the inclusion of NeXus you need the declaration of the NeXus API which is given in the file \texttt{napi.h}
 as well as the library \texttt{libNeXus} either in statically or dynamically linked form. For the HDF library you need the HDF API, which is
given in the file \texttt{mfhdf.h} and the library itself \texttt{libmfhdf}. With this knowledge you may tell your configuration system 
that it has to look for these files.

This has to be done in the \texttt{configure.ac} file. Please insert after line 9 the following lines:
\begin{verbatim}
AC_CHECK_HEADERS([napi.h mfhdf.h], 
        [], 
        AC_MSG_ERROR([Did not found the NeXus headers]))
\end{verbatim}
These  lines instruct the \texttt{autotools} to search in the standard include pathes for the files \texttt{napi.h} and \texttt{mfhdf.h}.
If one of these file was not found the \texttt{configure} script stops with an error message:
\begin{verbatim}
configure: error: Did not found the NeXus headers.
\end{verbatim}

For searching the HDF library you have to add these lines into your file \texttt{configure.ac}:
\begin{verbatim}
AC_CHECK_LIB(mfhdf, Hclose, 
        AC_MSG_RESULT([HDF library found]), 
        AC_MSG_ERROR([Did not found the HDF library]), 
	[-ldf -ljpeg -lz])
\end{verbatim}

It is designed to search for a special function (in this case \texttt{Hclose}) in the library \texttt{libmfhdf} to test the existence 
of the library. If it is found \texttt{configure} outputs: 
\begin{verbatim}
HDF library found.
\end{verbatim}
or else it gives an error message that it did not found the library and stops its execution.

For the search of the NeXus library you should add:
\begin{verbatim}
AC_CHECK_LIB(NeXus, nxiclose_, 
         [], 
         AC_MSG_ERROR([Did not found the NeXus library]))
\end{verbatim}


The \verb+AC_CHECK_LIB+ assumes that the tested function has no parameters. It
tests only the linking of the program, if you want to test the right parameters of a certain function you have to perform more 
sophisticated tests.  For this a deeper look into the manuals \cite{autoconf}, \cite{automake},\cite{libtool} 
or the book \cite{autotools} helps.


\section{Summary}
This paper gives a short introduction into the use of \texttt{autotools}, i.e. \texttt{autoconf}, \texttt{automake}, and \texttt{libtool}.
At hand of a little example its usage is explained, and it is discussed how the project may be extended and adopted to the specific 
environment. 
The use of autotools simplifies the creation and distribution of software which has to run on different systems.


\begin{thebibliography}{99}
\bibitem{autotools} G.V. Vaughan, B. Elliston, T. Tromey, and I.L. Taylor, \textsc{GNU AUTOCONF, AUTOMAKE, and LIBTOOL}, New Riders, 2001,
http://www.newriders.com/
\bibitem{autoconf} D. MacKenzie, B. Elliston, and A. Demaille, GNU autoconf manual
\bibitem{wwwautoconf} http://www.gnu.org/software/autoconf/
\bibitem{automake} D. MacKenzie, and T. Tromey, GNU Automake manual
\bibitem{wwwautomake} http://www.gnu.org/software/automake/
\bibitem{libtool}G. Matzigkeit, A. Oliva, Th. Tanner, and G.V. Vaughan. GNU libtool manual
\bibitem{wwwlibtool} http://www.gnu.org/software/libtool/
\bibitem{hdf}http://hdf.ncsa.uiuc.edu/
\bibitem{nexus}http://www.neutron.anl.gov/nexus/
\end{thebibliography}
\end{document}